\documentclass[aps,prb,twocolumn,groupedaddress,showpacs]{revtex4}%
\usepackage{epsfig}
\usepackage{amsmath}
\usepackage{amsfonts}
\usepackage{amssymb}
\usepackage{graphicx}%
\providecommand{\U}[1]{\protect\rule{.1in}{.1in}}
\begin{document}
\title{Structural and Electronic properties of the 2D Superconductor CuS with
1$\frac{1}{3}$-valent Copper}
\author{I.I. Mazin}
\affiliation{Code 6390, Naval Research Laboratory, Washington, DC 20375, USA}

\begin{abstract}
We present first principle calculations of the structural and electronic
properties of the CuS covellite material. Symmetry-lowering structural transition
is well reproduced. However, the microscopic origin of the transition is unclear.
The calculations firmly establish that the so far controversial Cu valency in
this compound is 1.33.
We also argue that recently reported high-temperature superconductivity in CuS
is unlikely to occur in the stoichiometric defect-free material, since the 
determined Cu valency is too close to 1 to ensure proximity to a Mott-Hubbard 
state and superexchange spin fluctuations of considerable strength.
On the other hand, one can imagine a related system with more  
holes per Cu in the same structural motif ($e.g.$, due to defects or O impurities)
in which case combination of superexchange and an enlarged compared to CuS Fermi
surface may lead to unconventional
superconductivity, similar to HTSC cuprate, but, unlike them, of an  $f$-wave symmetry. 

\end{abstract}

\maketitle

\begin{figure}[tbh]
\begin{center}
\centerline{\includegraphics*[width=0.95 \columnwidth]{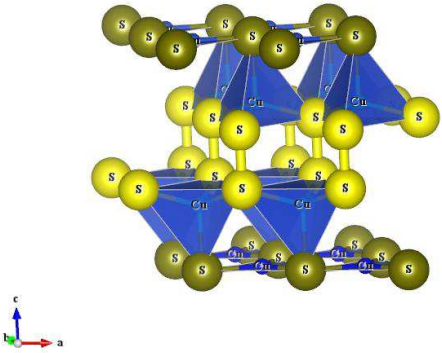}}
\end{center}
\caption{(color online) Crystal structure of covellite in the high-symmetry phase.
Large yellow spheres indicate sulfure, and blue spheres copper. 
The dark spheres form the planar CuS layers (Cu1), and the light spheres form
the warped Cu$_2$S$_2$ bilayers (Cu2). Fat yellow sticks indicate strong covalent
bonds inside the S2-S2 dumbells.
}
\label{str}
\end{figure}

Copper sulfide in the so-called covellite structure (Fig. \ref{str}) has recently attracted
attention due to a new report about possible superconductivity at 40
K\cite{Raveau}. This report has been met with understandable skepticism,
because previous researches\cite{sc1,sc2} reported reproducible
superconductivity at rather low temperatures, around 1-2 K. On the other hand,
inspection of the literature reveals that reported physical properties of 
covellite are drastically different in different papers. For instance, one
paper reported a well defined Curie-Weiss magnetic susceptibility\cite{1st},
while others observed a nearly constant behavior consistent with the Pauli
susceptibility in absence of any local moments.

Structural properties of the covellite are also intriguing. At room
temperature it consists of triangular layers of Cu and S, stacked as follows,
using standard hexagonal stacking notation: Cu1 and S1 form layers $A$ and $B$, at
the same height, so that Cu1 has coordination of three and no direct overlap.
Cu2 and S2 form layers $B$ and $C$, so that Cu2 is directly above S1 and bonds
with it, too, albeit more weakly than to S2. Thus, compared to the S2 layer,
the Cu2 layer is closer to
the Cu1+S1 one, and Cu2 appears to be inside a tetrahedron, closer to its
base. The next layer is again $C$, so that two S2 atoms are right on top of
each other and form a strongly covalent bond, the shortest bond in this
system, essentially making up an S$_{2}$ molecule.

At $T=55$ K the system spotaneously undergoes a transition from a hexagonal
structure to a lower symmetry orthorhombic structure. To a good approximation,
the transition amounts to sliding the Cu2-S2 plane with respect to the Cu1-S1
plane by 0.2 \AA , and the two neighboring Cu2-S2 planes by 0.1 \AA \ with
respect to each other, in the same direction. The bond lengths change very
little, one of the three Cu2-S2 bonds shortens by 0.04 \AA , and the S2-S2
bonds lengthens by 0.05 \AA , and all other bonds remain essentially
unchanged. Note that such transitions are quite uncommon for metals, but
rather characteristic of insulating Jahn-Teller systems. Transport properties
are hardly sensitive to this transition, which is however clearly seen in the
specific heat.

Thus there are three questions to be asked. First, what is the nature of the
low temperature symmetry-lowering? Second, why some experiments indicate pure
Pauli susceptibility, while others observe local moments (through Curie-Weiss
behavior)? Third, why one particular experiment sees indications of high
temperature superconductivity, while others do not? Of, course, there is
always a chance the ``outliers'' experiments are simply incorrect, but
it is always worth asking the question, whether some sample issues may

possibly account for such discrepancies.

In order to adress the first question, we have performed density functional
calculations (DFT) of both hexagonal (H) and orthorhombic (O) structure.
First, we optimized the crystal structure using the standard VASP program with
default settings (including gradient corrections),
 and starting from the experimental structure as reported in
Ref. \onlinecite{1st}. 

After that, all calculations in the determined crystal structures were performed using 
the standard all-electron LAPW code {\sc WIEN2k}. We have also verified that the 
calculated forces in the optimized structures are small enough. As a technical note,
to obtain full convergences in the energy differences we had to go up to $RK_{\max}=9$.

\begin{table}[tb]
\caption{The calculated total energy (meV/cell) of the low-temperature orthorhombic and the
high-temperature hexagonal structure, using either the experimental or the calculated
optimized parameters. Structural parameters used, as well as selected bond length (\AA)
are also shown. Note that one unit cell includes 6 formula units. The cell volume is given 
in $\AA ^3$
The last column corresponds to the orthorhombic structure with
internal coordinates optimized, while keeping the experimental
unit cell.
}
\begin{tabular}
[c]{|l|l|l|l|l|l|}
\hline
& H-exp & O-exp & H-calc & O-calc&O-c.o.\\ \hline
$a$ & 3.789 & 3.760 & 3.807 & 3.793&3.760\\
$b$ & 3.789 & 6.564 & 3.807 & 6.623&6.564\\
$c$ & 16.321 & 16.235 & 16.496 & 16.453&16.235\\
$z_{Cu2}$ & 0.1072 & 0.1070 & 0.1069 & 0.1077&0.1083\\
$z_{S2}$ & 0.0611 & 0.0627 & 0.0639 & 0.0646&0.0651\\
$y_{Cu1}$ & n/a & 0.6377&n/a  &0.6227 &0.6077\\
$y_{Cu2}$ & n/a &0.3372 &n/a  &0.3410 &0.3413\\
$y_{S1}$ & n/a &0.3068 &n/a  &0.2917 &0.2760\\
$y_{S2}$ &n/a & 0.0008  &n/a &0.0064&0.0069\\

Cu1-S1 & 3$\times$2.19 & 2$\times 2.18$ & 3$\times2.20$ & 2$\times 2.20$& 2$\times 2.20$\\
       &               &          2.17  &               &          2.19& 2.19\\
Cu2-S1 & 2.33 & 2.33 & 2.36 & 2.33&2.34\\
Cu2-S2 & 3$\times2.31$ & 2$\times 2.30$ & 3$\times2.31$ & 2$\times 2.30$& 2$\times 2.30$\\
 &  &2.28 &  & 2.33& 2.33\\
S2-S2 & 1.99 & 2.04 & 2.11 & 2.13&2.12\\
Volume&202.9&200.3&207.0&206.7&207.3\\
Energy & 0 & -85 & -258 & -265& -189 \\
\hline
\end{tabular}
\end{table}

The results are shown in Table 1. Even though there is some discrepancy
between the calculated and the experimental low-temperature structures
(mostly in terms of an overall overestimation of the equilibrium volume), 
 the
correct symmetry lowering is well reproduced. 
In fact, given that only one paper has reported internal positions for the
orthorhombic structre, and the same paper found a Curie-Weiss law, suggesting, as 
discussed below, crystallographic defects in their sample, it is fairly possible that
the calculations predict the structure of an ideal material better than this one
experiment has measured.
 
\begin{figure}[tbh]
\begin{center}
\centerline{\includegraphics*[width=0.95 \columnwidth]{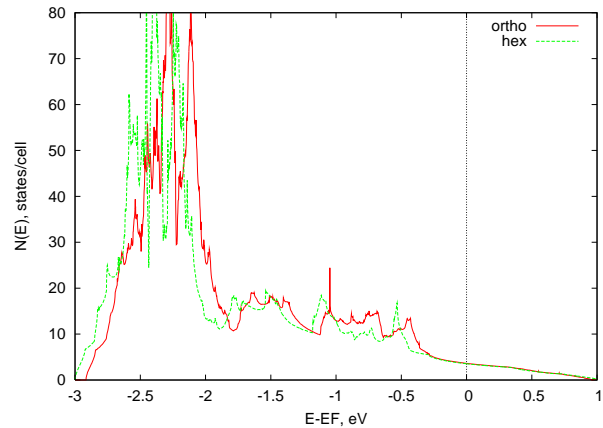}}
\end{center}
\caption{ (color online)
Calculated density of states in the high-temperature 
(``hex'') and low temperature (``ortho'') structures, using
in both cases optimized parameters.
}
\label{DOS}
\end{figure}

\begin{figure}[!t]
\begin{center}
\centerline{\includegraphics*[width=0.5\columnwidth, angle=60 ]{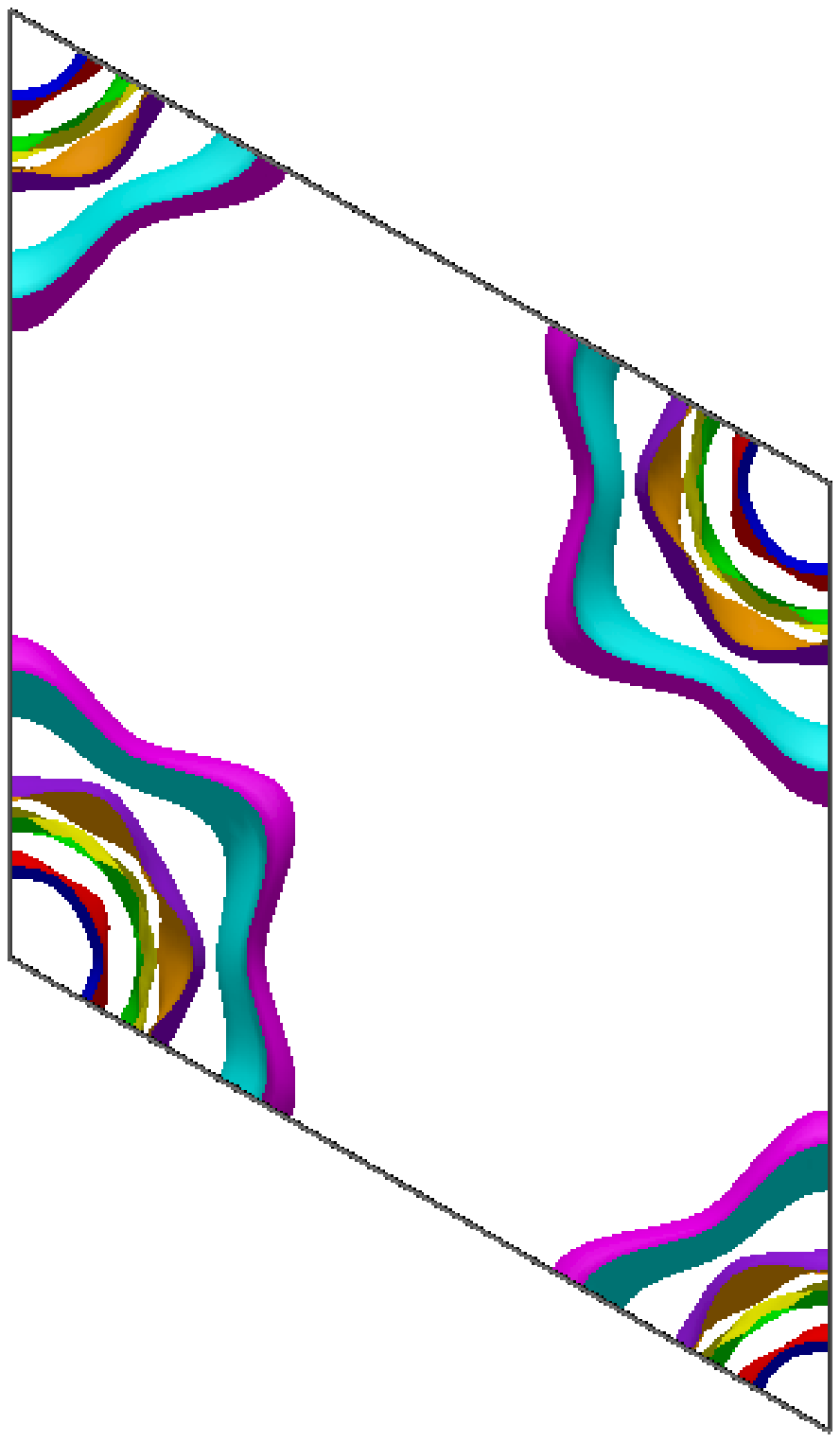}}
\vskip -10mm
\centerline{\includegraphics*[width=0.6\columnwidth, angle=90]{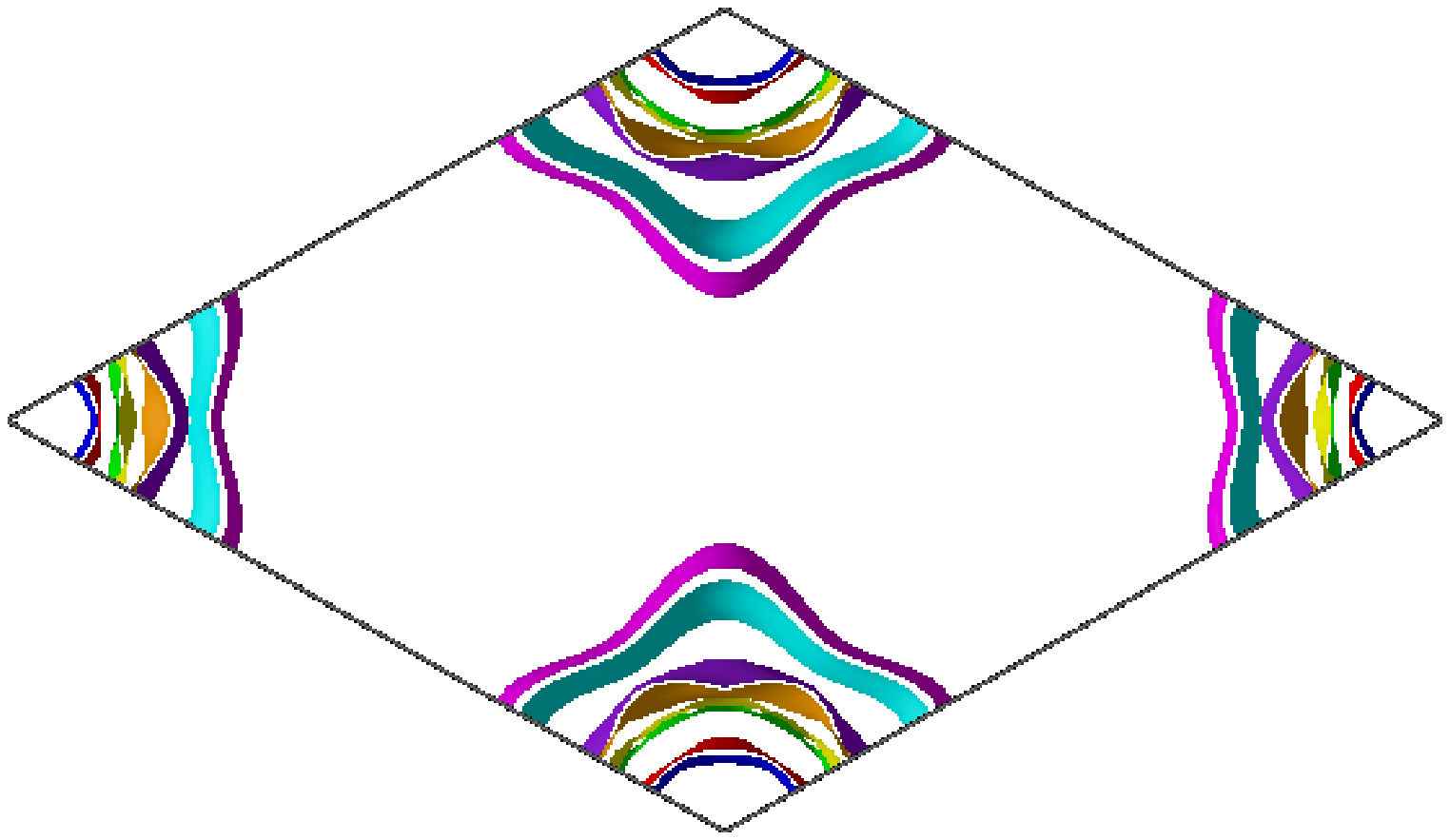}}
\end{center}
\caption{(color online)
Calculated Fermi surfaces in the hexagonal (top) and orthorhombic (bottom) structure,
viewed along the $c$-axis. Note reduced $k_z$ dispersion in the bottom panel.
}
\label{FS}
\end{figure}

A more important question now is, what 
the mechanism for this well-reproduce 
symmetry lowering can be? Ionic symmetry-lowering mechanisms (such as Jahn-Teller) 
are excluded in a
wide band metal like CuS. Typically, a lower symmetry is stabilized in a metal
if it results in a reduced density of states at the Fermi level (``quasinesting mechanism''). However, the
density of states at the Fermi level does not change at this transition (Fig.
\ref{DOS}), and the states below Fermi actually shift slightly upward. Thus,
one-electron energy is not the reason for the transition. A look at the
calculated Fermi surfaces (Fig. \ref{FS}) shows that while they become more 2D
in the orthorhombic structure (the in-plane plasma frequency remains the same,
$\approx4.0$ eV, while that out of plane, 1.36 eV, drops by 12\%), there is no
shrinkage in their size.

We cannot say definitively what causes the low-temperature symmetry lowering
in CuS, but we can say confidently that it is not van der Waals interaction as
conjectured in Ref. \onlinecite{W} (for one, it would not be reproduced in
LGA/GGA calculations with requested accuracy, and, also, as dicussed below,
the interplanar bonding is covalent, and not van der Waals), and not a typical metallic mechanism driven by
Fermi surface changes. One candidate is ionic Coulomb interaction. Indeed the
calculated Ewald energy is noticeably lower in the orthorhombic structure. However
the Ewald energy is only part of total electrostatic energy, so from this fact alone
one cannot derive definitive conclusions.
\begin{figure}[tbh]
\begin{center}
\centerline{\includegraphics*[width=0.95 \columnwidth]{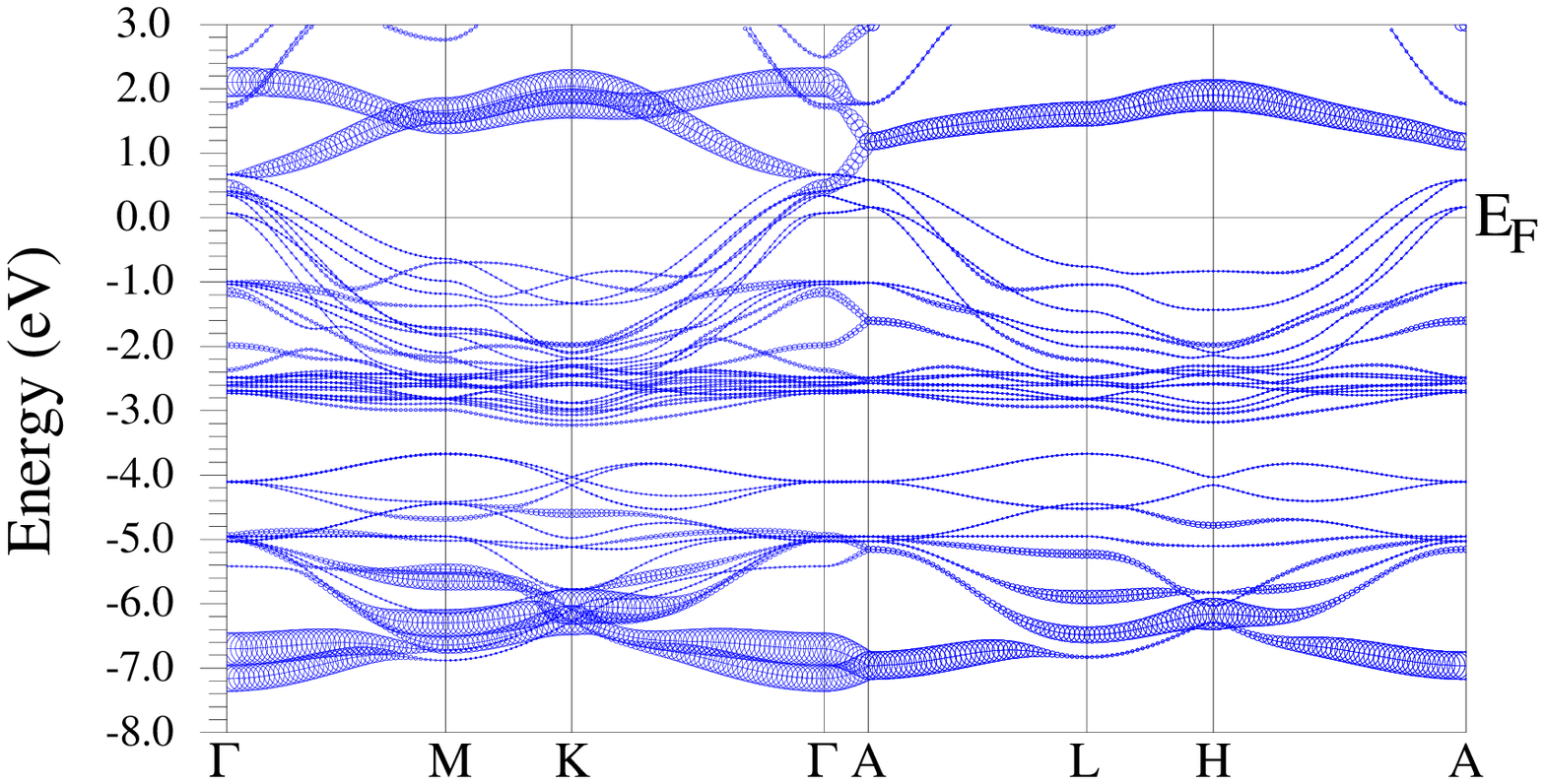}}
\centerline{\includegraphics*[width=0.95 \columnwidth]{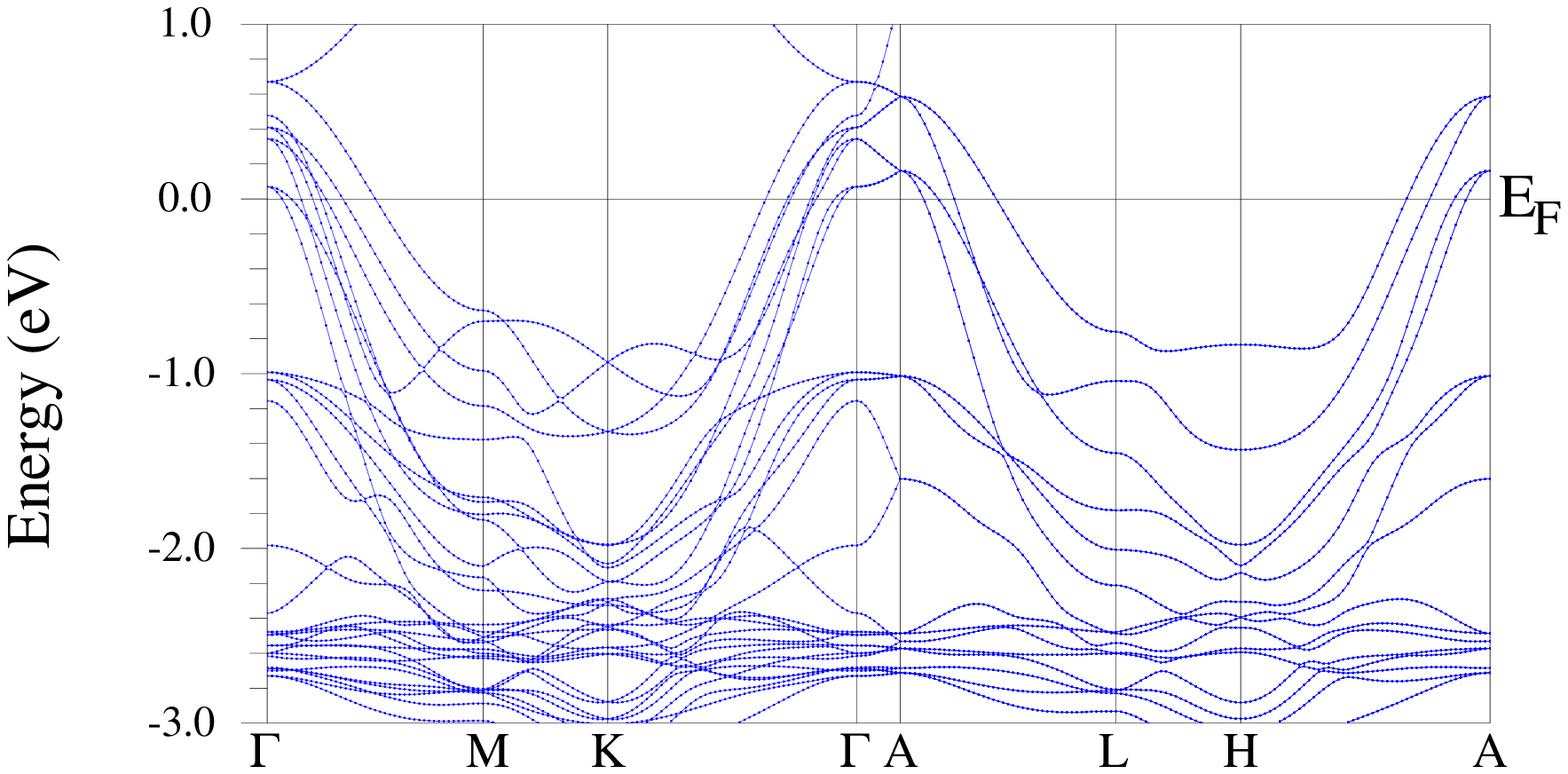}}
\end{center}
\caption{
Calculated band dispersions in the hexagonal  
structure. The points $\Gamma$, M and K are in the central plane ($k_z=0$) and
A, L and H in the basal plane ($k_z=\pi/c$). In the top panel, the width 
of the lines is proportional to the amount of the S2-$p_z$ character in the corresponding
states.
}
\label{bands}
\end{figure}

Let us now discuss the electronic structure. Since the differences between the
two structures are very small, we shall limit our discussion by the
high-temperature hexagonal structure. The calculated band structure is shown
in Fig. \ref{bands}. Note two sets of bands, one at -7 eV and the other at 1
eV, of strong S2-$p_{z}$ character. These are bonding and antibonding bands
of the S2-S2 dumbells. Historically, there has been a heated discussion of
the Cu valency in this compound, and what is an appropriate ionic model. Both
(Cu$^{+1})_{3}$(S$_{2}^{-2})($S$^{-1})$ (Ref. \onlinecite{1st}) and (Cu$^{+1})_{3}%
($S$_{2}^{-1})($S$^{-2})$ (Ref. \onlinecite{W}) have been discussed, assuming
monovalent copper. On the other hand, XPS\cite{XPS} and NQR\cite{Irek} data
indicated Cu valency larger than 1, but smaller than 1.5. From our calculations
it is immediately obvious that S1 is divalent, while S2 is monovalent (the
antibonding $p_{z}$ band of the S2-S2 dimer is 1 eV above the Fermi level, while
all S1-derived bands are below the Fermi level), so that Cu has
valency 1.33, and the appropriate ionic model is (Cu$^{+4/3})_{3}$(S$_{2}%
^{-2})($S$^{-2}).$

This means that the Cu $d-$band has 1/3 hole per Cu ion, 2.5 times fewer than
in the high-T$_{c}$ cuprates (optimal doping corresponds to 0.8-0.85 holes).
This may be too far from half filling for strong correlation effects, but it
is nevertheless suggestive of possible spin fluctuations. We will return to
this point later.

In order to understand the Cu $d$ bands near the Fermi level, let us consider
a simple tight binding model with two $d$ orbitals, with $m=\pm2$
(corresponding to combinations of the $x^{2}-y^{2}$ and $xy$ cubic harmonics, which belong to the
same representation in the hexagonal group). Since these orbitals are the ones
spread most far in the plane, their hybridization with S is the strongest and they
form the highest antibonding states near the $\Gamma$ point, crossing the
Fermi level. Integrating out the S $p_{x,y}$ orbitals we arrive at the
following model band structure:%
\begin{widetext}
\begin{equation}
E_{k}=\frac{1}{2(\varepsilon_{d}-\varepsilon_{p})}\left[  (3t_{pd\sigma}%
^{2}+4t_{pd\pi}^{2})\sum_{i}\cos(\mathbf{k\cdot R}_{i})\pm(3t_{pd\sigma}%
^{2}-4t_{pd\pi}^{2})\sqrt{\sum_{i}\cos^{2}(\mathbf{k\cdot R}_{i})-\sum
_{i>j}\cos(\mathbf{k\cdot R}_{i})\cos(\mathbf{k\cdot R}_{j})}\right],
\label{dispersion}%
\end{equation}
\end{widetext}
where $t$ are the Cu-S hopping amplitudes, and \textbf{R}$_{i}$ are the three
standard triangular lattice vectors, $\sum_{i}\mathbf{R}_{i}=0.$ Note that these
bands are degenerate at $\Gamma,$ unless spin-orbit coupling is taken into
account. Near the top of the band the dispersion is isotropic, and away from
it the Fermi surface develops a characteristic hexagonal rosette shape (Fig.
\ref{FS}).

Let us now look at the calculated bands (Fig. \ref{bands}). It is more
instructive to concentrate on the righ hand side of Fig. \ref{bands}, where
the $k_{z}$ dispersion does not obscure the states degeneracy. We see, as predicted by the
model, three sets of  nearly parabolic bands, each four times degenerate at the
point A=(0,0,$\pi/c).$ One of them is below the Fermi level and two above,
forming the eight FS sheets we see in Fig. \ref{FS}. The middle bands are
predominantly formed by the Cu1 and the lower (fully occupied) and upper ones
by the Cu2, although there is substantial mixture of all three Cu orbitals.
The average occupation of Cu d orbitals, as described above, is 1/3 hole per
Cu, too small to form a magnetic state, even in LDA+U with $U\sim8$ eV (as
verified by direct calculations). Formation of an ordered magnetic state is
additionally hindered by the fact that supexchange is this case is
antiferromagnetic, and frustrated, as it should be on a triangular lattice.
One may think that additional hole doping, achieved through Cu vacancies,
broken S-S bonds or interstitial oxygen (note that this structures includes
large pores, one per formula unit, in each Cu-S layer) should bring the
$d-$bands closer to half-occupancy and promote local magnetic moments. Note
that in at least one experimental paper a Curie-Weiss behavior was reported,
corresponding to 0.28 $\mu_{B}/$Cu\cite{1st}, and in another a weak, but
inconsistent with the Pauli law, temperature dependence was found\cite{NMR}, while
other authors reported temperature-independent susceptibility.
\begin{figure}[tbh]
\begin{center}
\centerline{\includegraphics*[width=0.7\columnwidth, angle=0 ]{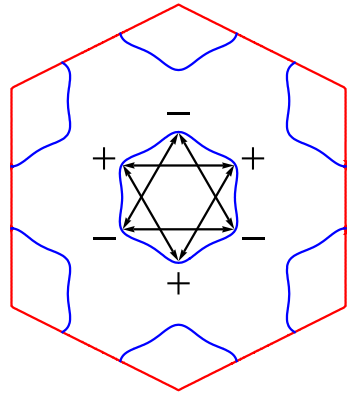}}
\end{center}
\caption{(color online)
A model Fermi surface, calculated using Eq. \protect\ref{dispersion}, overlapped with the wave vectors 
corresponding to superexchange on a triangular lattice. The signs show a possible
$f-$wave pairing state, consistent with superexchnage-induced spin fluctuations.
}
\label{model}
\end{figure}

One can speculate that the unexpected high-temperature superconductivity
observed by Raveau $et$ $al$\cite{Raveau} is a phenomenon of the same sort,
namely that this superconductivity forms in a portion of a sample, the same
portion where some previous researchers observed local magnetic moments. As
discussed above, it is highly unlikely that a stoichiometric, defectless CuS
sample would support either local moments or unconventional superconductivity.
However, it is of interest to consider a hypothetical situation that would
take place if such moments were present. Indeed in that case one can write
down the superexchange interaction between the nearest neighbors as
antiferromagnetic Heisenberg exchange, in which case in the reciprocal space
it will have the following functional form:%
\begin{equation}
J(\mathbf{q)=}J\sum_{i}\cos(\mathbf{k\cdot R}_{i}).\label{J}%
\end{equation}

In Fig. \ref{model} we show an example of a Fermi surface generated for the
model band structure (Eq. \ref{dispersion}), for the simplest case of $t_{pd\pi
}=0$. The wave vectors corresponding to the peaks of the superexchange
interaction (\ref{J}) are shown by arrows. An interesting observation is that
for this particular doping this superexchange interaction (or, better to say,
spin fluctuations generated by this superexchange) would be pairing for a
triplet $f-$state shown in the same picture\cite{f}. Indeed, the superexchange vectors
always span the lobes of the order parameter with the same sign. Since in a
triplet case spin-fluctuations generate an attractive interaction, it will be
pairing for the geometry shown in Fig.  \ref{model}. Note that this is
opposite to high-$T_{c}$ cuprates, where the superexchange interaction spans
parts with the opposite parts of the $d-$wave order parameter, but in a
singlet channel this interaction is repulsive, and therefore pairing when the
corresponding parts of the Fermi surface have opposite signs of the order parameter.

While the model Fermi surface shown in Fig. \ref{model} is roughly similar to that
calculated in the stoichiometric CuS, the system at this doping is too far
from the ordered magnetism to let us assume sizeable superexchange-like
magnetic fluctuations. As mentioned,  our attempts to stabilize an antiferromagnetic
(more precisely, ferrimagnetic, since we only tried collinear magnetic
patterns) using a triple unit cell failed, even in LDA+U. One may
think of a hole doped system, where superexchange is operative and the inner
Fermi surfaces (albeit not the outer ones) have geometry similar to that
featured in Fig. \ref{model}. 

Of course, it may not be possible to stabilize a system 
at sufficient hole doping and retain the required crystallography. We prefer to think 
about the model discussed in the perevious paragraphs as {\it inspired} by the 
CuS covellite, bit not necessary applicable to actual materials derived from this one.
The reason we paid so much attention to it is that this is a simple generic model, 
describing any triangular planar structure with transition metals and ligands in the same plane, as
in the covellite, in case where correlations are sufficiently strong to bring about 
spin fluctuations controlled by superexchange. It is quite exciting that, compared
to the popular spin-fluctuation scenario of superconductivity in cuprates, to which it 
is conceptually so similar, this simple generic model results in a completely 
different superconducting state, triplet $f$, as opposed to singlet $d$.
This finding may have implications far beyound this particular material and
(yet unconfirmed) superconductivity in it.


\begin{thebibliography}{99}
\bibitem{f} To be specific, this states is $D_{6h}(\Gamma^-_4)$, using  the notations 
of Sigrist and Ueda\protect\cite{SU}, with the basis functions 
${\bf \hat{z}}(k_y(k_y^2-3k_x^2)$ or $k_z[(k_y^2-k_x^2){\bf \hat{y}}-2k_xk_y{\bf \hat{x}}]$.
\end{thebibliography}
\end{document}